\documentclass{iau} 
\usepackage{natbib}
\usepackage{graphicx,color}
\usepackage[T1]{fontenc}
\usepackage[utf8]{inputenc}
\usepackage{lmodern}

\usepackage[colorlinks,bookmarks]{hyperref}
\definecolor{linkblue}{rgb}{0.0,0.0,0.6}
\hypersetup{linkcolor=linkblue, citecolor=linkblue, urlcolor=linkblue}

\newcommand{\natastron}{{Nat. Astron.}}
\newcommand{\aj}{AJ}

\newcommand{\mnras}{MNRAS}

\newcommand{\apj}{ApJ}

\title[The RAR emergent nature] 
{The radial acceleration relation and its emergent nature}

\author[D.C.~Rodrigues \& V.~Marra]{ 
Davi C. Rodrigues
\and Valerio Marra}
\affiliation{Núcleo de Astrofísica e Cosmologia, PPGCosmo \& Departamento de Física,\\ Universidade Federal do Espírito Santo, 29075-910, ES, Brazil.  \\ emails: {\tt  davi.rodrigues@cosmo-ufes.org, marra@cosmo-ufes.org}}

\pubyear{2020}
\volume{359}  
\jname{Galaxy evolution and feedback across different environments}
\editors{T. Storchi-Bergmann, R. Overzier, W. Forman \& R. Riffel, eds.}

\begin{document}

\maketitle

\begin{abstract}
We review  some of our recent results about the Radial Acceleration Relation (RAR) and its interpretation as either a fundamental or an emergent law. The former interpretation is in agreement with a class of modified gravity theories that dismiss the need for dark matter in galaxies (MOND in particular). Our most recent analysis, which includes refinements on the priors and the Bayesian test for compatibility between the posteriors,  confirms that the hypothesis of a fundamental RAR is rejected at more than 5$\sigma$ from the very same data that was used to infer the RAR.


\end{abstract}

\firstsection

\vspace*{-0.3cm}
\section{Introduction}

The Radial Acceleration Relation (RAR) \citep{McGaugh:2016leg} shows a sharp correlation between two accelerations associated to galaxy rotation curves. Since this correlation with its small dispersion is not an obvious outcome of the standard dark matter picture, several works interpreted  the RAR as  evidence for modified gravity such as the Modified Newtonian Dynamics (MOND) \citep[e.g.,][]{Li:2018tdo}. For the latter model, such correlation is a fundamental property of gravity, which is achieved by introducing a fundamental acceleration scale $a_0$, while removing dark matter.

 \cite{Rodrigues:2018duc} have shown, using Bayesian inference and the SPARC data \citep{2016AJ....152..157L}, that the $a_0$ credible intervals for different galaxies are not compatible among themselves at more than $10\sigma$. Hence, also considering that high-quality rotation curve data were used, this led to a re-interpretation of the RAR as the strongest evidence against MOND as a gravitational theory \citep{Marra:2020sts}. Here we consider the approach of \cite{Rodrigues:2018duc} together with further recent refinements by \cite{Marra:2020sts} on the statistical analysis.

To evaluate if MOND works as a dark matter replacement in galaxies, one has to address the issue of finding $a_0$. A common practice is to fit many galaxies and take the median of the best-fit values. Doing so is not optimal since it neglects the information from the $a_0$ posterior distributions of the individual galaxies (i.e., the ``errors'' on $a_0$ for each one of the galaxies). Moreover, and most importantly, from those posteriors one can test if the observational data are compatible with the existence of a common $a_0$ value. If they are not compatible, then $a_0$ is not fundamental and the RAR is necessarily an emergent correlation. Assuming standard dark matter, the RAR must be emergent \citep[see e.g.,][]{2019ApJ...882....6S}. If the RAR is emergent it  can be useful (like many other emergent correlations), but it cannot directly reflect a fundamental property of gravity.

\vspace*{-0.5cm}

\section{Methods and results}

\begin{table}
\caption{The rejection level of the fundamental $a_0$ hypothesis.
}
\centering
\renewcommand{\arraystretch}{1.5}
\setlength{\tabcolsep}{20pt}
\begin{tabular}{lp{2 cm}p{3cm}}
\hline
Method & RAR sample &  ${\cal S}_2$ subsample  \\
&  (153 galaxies) &  (91 galaxies) \\
\hline 
Monte Carlo $X^2$ test   & $>5.7\sigma$ & $5.3\sigma$ \\
\hline
\end{tabular}
\label{tab:res}
\end{table}%

\begin{figure}
  \includegraphics[width=\textwidth]{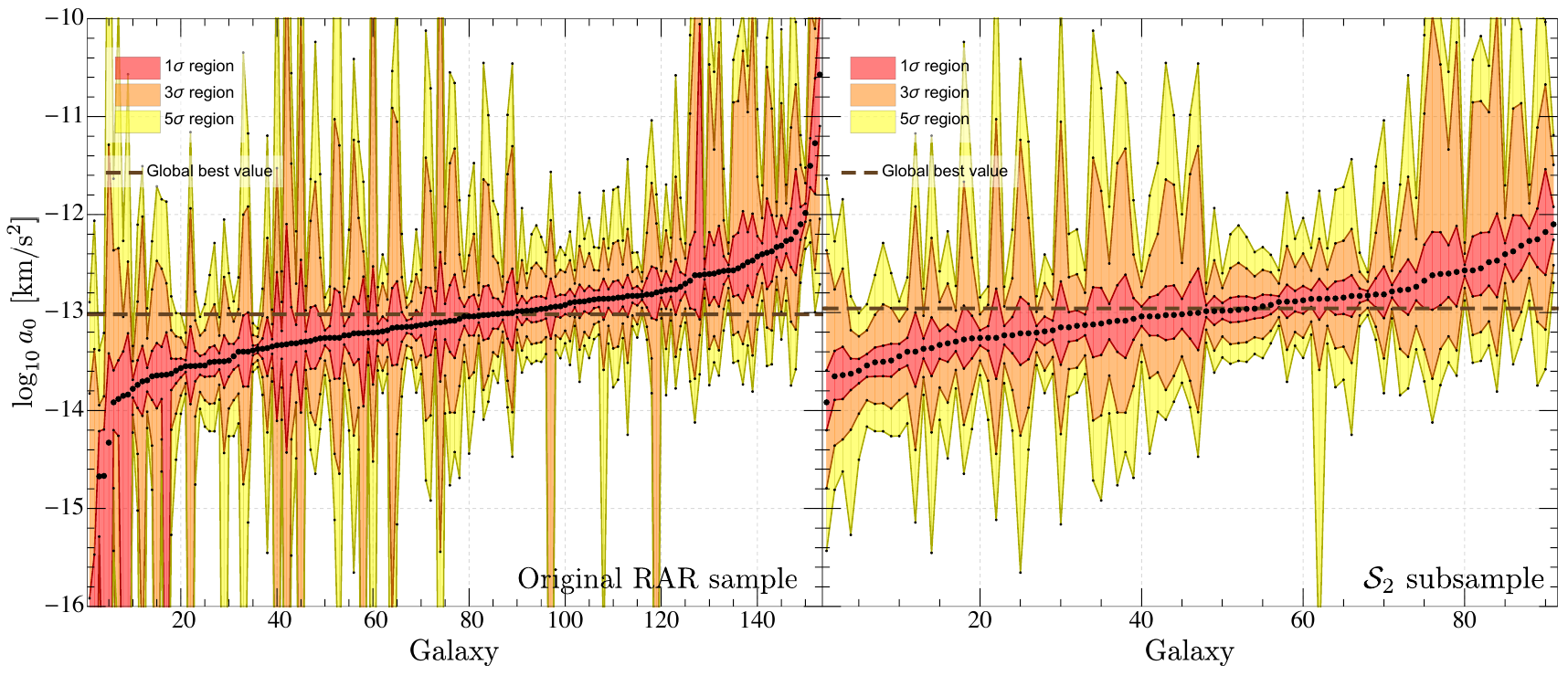}
  \caption{The $a_0$ modes (black dots) and the $1\sigma$, $3\sigma$ and $5\sigma$ credible intervals for each one of the galaxies.
  The left panel shows all the 153 RAR galaxies, while the right panel shows a subsample (${\cal S}_2$) with a stronger quality cut: galaxies with too high $\chi^2_{\rm min}$ values are also eliminated (91 galaxies are left). See \citet{Marra:2020sts} for further details.
  }
\end{figure}

\citet{Marra:2020sts} improved the methodology of \cite{Rodrigues:2018duc} by considering priors more closely related to the observational uncertainties \citep[see also][]{Rodrigues:2018lvw}. Beyond  the variation of $a_0$ from galaxy to galaxy, it is also considered the variation of mass-to-light ratios, distances and inclinations. Another improvement concerns the comparison of the credible intervals, since the method no longer employs Gaussian approximations \citep[see also][]{2020NatAs...4..132C,Rodrigues:2020squ}.
In particular, the  existence of a common $a_0$ value is tested using the $X^2$ statistics, an extension of the tension estimator of \citet{Verde:2013wza}, which is based on the Bayes factor.

We conclude with our opinion on MOND: historically it has stimulated relevant developments for galaxy astrophysics. As an effective theory for galaxy dynamics valid on average, it is useful and it is the RAR (i.e., a correlation between accelerations). As a theory for gravity, it has many problems, even for galaxy rotation curves.\\[-.2cm]

\noindent
{\small {\bf Acknowledgements.} DCR thanks the organizers of the GALFEED Symposium. DCR and VM also thank CNPq (Brazil) and FAPES (Brazil) for partial financial support.}


\vspace*{-0.4cm}

\begin{thebibliography}{}
\makeatletter
\relax
\def\mn@urlcharsother{\let\do\@makeother \do\$\do\&\do\#\do\^\do\_\do\%\do\~}
\def\mn@doi{\begingroup\mn@urlcharsother \@ifnextchar [ {\mn@doi@}
  {\mn@doi@[]}}
\def\mn@doi@[#1]#2{\def\@tempa{#1}\ifx\@tempa\@empty \href
  {http://dx.doi.org/#2} {doi:#2}\else \href {http://dx.doi.org/#2} {#1}\fi
  \endgroup}
\def\mn@eprint#1#2{\mn@eprint@#1:#2::\@nil}
\def\mn@eprint@arXiv#1{\href {http://arxiv.org/abs/#1} {{\tt arXiv:#1}}}
\def\mn@eprint@dblp#1{\href {http://dblp.uni-trier.de/rec/bibtex/#1.xml}
  {dblp:#1}}
\def\mn@eprint@#1:#2:#3:#4\@nil{\def\@tempa {#1}\def\@tempb {#2}\def\@tempc
  {#3}\ifx \@tempc \@empty \let \@tempc \@tempb \let \@tempb \@tempa \fi \ifx
  \@tempb \@empty \def\@tempb {arXiv}\fi \@ifundefined
  {mn@eprint@\@tempb}{\@tempb:\@tempc}{\expandafter \expandafter \csname
  mn@eprint@\@tempb\endcsname \expandafter{\@tempc}}}

\bibitem[\protect\citeauthoryear{{Cameron}, {Angus}  \& {Burgess}}{{Cameron}
  et~al.}{2020}]{2020NatAs...4..132C}
{Cameron} E.,  {Angus} G.~W.,   {Burgess} J.~M.,  2020, \mn@doi [\natastron]
  {10.1038/s41550-019-0998-2}, 4, 132.

\bibitem[\protect\citeauthoryear{{Lelli}, {McGaugh}  \& {Schombert}}{{Lelli}
  et~al.}{2016}]{2016AJ....152..157L}
{Lelli} F.,  {McGaugh} S.~S.,   {Schombert} J.~M.,  2016, \mn@doi [\aj]
  {10.3847/0004-6256/152/6/157}, 152, 157,
  [\href{https://arxiv.org/abs/1606.09251}{1606.09251}].

\bibitem[\protect\citeauthoryear{Li, Lelli, McGaugh  \& Schormbert}{Li
  et~al.}{2018}]{Li:2018tdo}
Li P.,  Lelli F.,  McGaugh S.,   Schormbert J.,  2018, \mn@doi [Astron.
  Astrophys.] {10.1051/0004-6361/201732547}, 615, A3,
  [\href{https://arxiv.org/abs/1803.00022}{1803.00022}].

\bibitem[\protect\citeauthoryear{Marra, Rodrigues  \& de Almeida}{Marra
  et~al.}{2020}]{Marra:2020sts}
Marra V.,  Rodrigues D.~C.,   de Almeida {\'A}.~O.,  2020, \mn@doi [\mnras]
  {10.1093/mnras/staa890}, 494, 2875,
  [\href{https://arxiv.org/abs/2002.03946}{2002.03946}].

\bibitem[\protect\citeauthoryear{McGaugh, Lelli  \& Schombert}{McGaugh
  et~al.}{2016}]{McGaugh:2016leg}
McGaugh S.,  Lelli F.,   Schombert J.,  2016, \mn@doi [Phys. Rev. Lett.]
  {10.1103/PhysRevLett.117.201101}, 117, 201101,
  [\href{https://arxiv.org/abs/1609.05917}{1609.05917}].

\bibitem[\protect\citeauthoryear{Rodrigues, Marra, Del~Popolo  \&
  Davari}{Rodrigues et~al.}{2018a}]{Rodrigues:2018duc}
Rodrigues D.~C.,  Marra V.,  Del~Popolo A.,   Davari Z.,  2018a, \mn@doi
  [\natastron] {10.1038/s41550-018-0498-9}, 2, 668,
  [\href{https://arxiv.org/abs/1806.06803}{1806.06803}].

\bibitem[\protect\citeauthoryear{Rodrigues, Marra, Del~Popolo  \&
  Davari}{Rodrigues et~al.}{2018b}]{Rodrigues:2018lvw}
Rodrigues D.~C.,  Marra V.,  Del~Popolo A.,   Davari Z.,  2018b, \mn@doi
  [\natastron] {10.1038/s41550-018-0614-x}, 2, 927,
  [\href{https://arxiv.org/abs/1811.05882}{1811.05882}].

\bibitem[\protect\citeauthoryear{Rodrigues, Marra, Del~Popolo  \&
  Davari}{Rodrigues et~al.}{2020}]{Rodrigues:2020squ}
Rodrigues D.~C.,  Marra V.,  Del~Popolo A.,   Davari Z.,  2020, \mn@doi
  [\natastron] {10.1038/s41550-019-0999-1}, 4, 134,
  [\href{https://arxiv.org/abs/2002.01970}{2002.01970}].

\bibitem[\protect\citeauthoryear{{Stone} \& {Courteau}}{{Stone} \&
  {Courteau}}{2019}]{2019ApJ...882....6S}
{Stone} C.,  {Courteau} S.,  2019, \mn@doi [\apj] {10.3847/1538-4357/ab3126},
  882, 6, [\href{https://arxiv.org/abs/1908.06105}{1908.06105}].

\bibitem[\protect\citeauthoryear{Verde, Protopapas  \& Jimenez}{Verde
  et~al.}{2013}]{Verde:2013wza}
Verde L.,  Protopapas P.,   Jimenez R.,  2013, \mn@doi [Phys. Dark Univ.]
  {10.1016/j.dark.2013.09.002}, 2, 166,
  [\href{https://arxiv.org/abs/1306.6766}{1306.6766}].

\makeatother
\end{thebibliography}
\end{document}